\documentclass{jfm}

\usepackage{graphicx}
\usepackage{natbib}
\usepackage{amsmath}
\usepackage{amssymb}
\usepackage{xcolor}

\renewcommand\Re{\mbox{Re}} 
\newcommand\Rey{\mbox{\textit{Re}}}  

\newcommand{\av}[1]{\overline{#1}}
\newcommand{\fl}[1]{{#1}'}

\renewcommand{\d}{\textnormal{d}}

\newcommand{\avS}[1]{\left< {#1} \right>_S}
\newcommand{\avV}[1]{\left< {#1} \right>_V}
\renewcommand{\vec}[1]{\boldsymbol{#1}}
\newcommand{\del}[0]{\boldsymbol{\nabla}}
\newcommand{\rev}[1]{#1}
\newcommand{\ignore}[1]{}

                      \newcommand{\CaptionCircleOpenSml}{{\small $\circ$}}                   
                    
\newcommand{\CaptionUpTriOpen}{{\large $\vartriangle$}}                
\newcommand{\CaptionDownTriOpen}{{\large $\triangledown$}}                

\newcommand{\CaptionSquareOpen}{{\small $\Box$}}

\newcommand{\CaptionDiamondOpen}{{\small $\Diamond$}}

\title{The turbulence boundary of a temporal jet}

\author[M. van Reeuwijk and M. Holzner]%
{M\ls A\ls A\ls R\ls T\ls E\ls N\ns van\ns R\ls E\ls E\ls U\ls W\ls I\ls J\ls K$^1$%
  \thanks{Email address for correspondence: m.vanreeuwijk@imperial.ac.uk}\ns
\\ and \\ M\ls A\ls R\ls K\ls U\ls S \ns H\ls O\ls L\ls Z\ls N\ls E\ls R$^2$}

\affiliation{$^1$Department of Civil and Environmental Engineering, Imperial College London, London, UK\\
$^2$Institute of Environmental Engineering, ETH Z\"{u}rich, Switzerland}

\date{\today}

\begin{document}
\maketitle

\begin{abstract}
\rev{We examine the structure of the turbulence boundary of a temporal plane jet at $\Rey=5000$ using statistics conditioned on the  enstrophy.
The data is obtained by direct numerical simulation and threshold values span 24 orders of magnitude, ranging from essentially irrotational fluid outside the jet to fully turbulent fluid in the jet core.
We use two independent estimators for the local entrainment velocity $v_n$ based on the enstrophy budget.
The data show clear evidence for the existence of a viscous superlayer (VSL) that envelopes the turbulence. The VSL is a nearly one-dimensional layer with low surface curvature. We find that both its area and viscous transport velocity adjust to the imposed rate of entrainment so that the integral entrainment flux is independent of threshold, although low-Reynolds-number effects play a role for the case under consideration. This threshold independence is consistent with the inviscid nature of the integral rate of entrainment. A theoretical model of the VSL is developed that is in reasonably good agreement with the data and predicts that the contribution of viscous transport and dissipation to interface propagation have magnitude $2 v_n$ and $-v_n$, respectively.
We further identify a turbulent core region (TC) and a buffer region (BR) connecting the VSL and the TC. 
The BR grows in time and inviscid enstrophy production is important in this region.
The BR shows many similarities with the turbulent-nonturbulent interface (TNTI), although the TNTI seems to extend into the TC.
The average distance between the TC and the VSL, i.e.\ the BR thickness is about 10 Kolmogorov length scales or half a Taylor length scale, indicating that intense turbulent flow regions and viscosity-dominated regions are in close proximity.}
\end{abstract}

\section{Introduction}
\label{sec:introduction}
Turbulent entrainment is the incorporation of ambient fluid at the boundary of turbulent flows such as free shear flows or at the free stream edge of turbulent boundary layers. It is an important process in a variety of engineering and geophysical flows controlling the turbulent transfer of mass, heat and momentum \citep{DaSilva2013,Pope2000, Stull1998, Thorpe2005}. A relevant, yet unresolved issue that has received renewed interest in recent years is the connection between processes that are associated with the large scale organization of the flow and processes that occur at the scale of the smallest eddies \citep[e.g.][]{Westerweel2005,DaSilva2010,Hunt2011,Philip2012,Wolf2012}.

The integral rate at which ambient fluid is incorporated into the turbulent flow, in the following referred to as \rev{global} entrainment, is independent of the small scale details of the flow, i.e. it does not depend on the viscosity or the energy dissipation mechanism.
The common entrainment assumption is that the global entrainment velocity $u_e$ is proportional to the typical velocity $\hat{u}$ inside the turbulent zone \citep{Morton1956, Turner1986}, usually the centreline velocity.
The entrainment coefficient $\alpha= u_e / \hat{u}$ is typically $O(0.1)$, but the value is far from universal and depends on the choice of the typical length scale $b$, the assumed shape of the velocity profile and can also depend on the initial conditions \citep[e.g.][]{Redford2012}.

Conversely, Corrsin emphasised a \rev{microscale} perspective, \rev{in the following referred to as local entrainment}, and suggested that the turbulence boundary is demarcated by a very thin viscosity-dominated laminar superlayer, whose local propagation velocity $v_n$ towards the non-turbulent region is determined by two parameters: the kinematic viscosity $\nu$ and the rate of dissipation of kinetic energy $\varepsilon$~\citep{Corrsin1955}. Consequently, $v_n \propto u_\eta$ where $u_\eta$ is the Kolmogorov velocity scale. The ratio of local entrainment velocity $v_n$ to global entrainment velocity $u_e$ is given by
\begin{equation}
  \frac{v_n}{u_e} \propto \frac{u_\eta}{\hat{u}} = \Re^{-1/4},
\end{equation}
and the Reynolds number dependence begs the question in which way the two views - local and global - are consistent.
The dependence on $\Re$ seems to imply that both surface area and viscous diffusion adjust to the imposed global entrainment rate such that the small scale details of how the vorticity is transferred are somehow forgotten across interactions of eddies with a large hierarchy of sizes~\citep{Townsend1976}.
By denoting the integral entrainment flux as $Q_e$, the global perspective suggests that $Q_e = u_e A$, where $A$ is the surface area based on the average distance of the turbulence interface to the core of the turbulent zone.
From the local perspective $Q_e = v_n S$ where $S$ is the total surface area of the contorted turbulence boundary. Equating the two expressions for $Q_e$ results in $u_e/v_n = S/A$ and therefore
\begin{equation}
\frac{S}{A} \propto Re^{1/4}.
\end{equation}
This means that $S$ must be large to compensate a slow viscous transfer of vorticity and to cancel out the viscosity dependence~\citep[e.g.][]{Tritton1988,Sreenivasan1989}.

Probably the simplest setting where turbulence propagates into non-turbulent fluid is the case without any mean flow, which can be realized via oscillating grid experiments, e.g.~\cite{Holzner2007,Holzner2008,Holzner2011}.
The results obtained in such a flow showed evidence for the presence of a laminar superlayer at the boundary of turbulent flow regions.
In particular, the analysis supported that $S$ is indeed given by a strongly convoluted surface and accounts for a large entrainment flux with a small characteristic velocity comparable to the Kolmogorov velocity~\citep{Holzner2011}.
A similar picture, i.e. $v_n\sim u_{\eta}$, recently emerged from the experiments in a round jet of \citet{Wolf2012}.
In all the experiments and simulations, the probability density functions (PDFs) of the entrainment velocity indicated that there is a large variation in entrainment velocities. In this context the term laminar superlayer is unfortunate as it suggests that the flow is layered without notable fluctuations.
Therefore, this layer will be termed the \emph{viscous} superlayer (VSL) in the remainder of the paper.

A somewhat different view emerges from direct numerical simulations of plane jets \citep{DaSilva2008,DaSilva2010,Taveira2013}, a plane wake \citep{Bisset2002} and experiments in a round jet \citep{Westerweel2005,Westerweel2009}, which focused on properties of the turbulent/nonturbulent interface (TNTI).
The TNTI seems to be thicker than the VSL predicted by Corrsin; that is, the thickness of the TNTI is comparable to the Taylor length scale $\lambda$, rather than than the Kolmogorov length scale $\eta$.
\rev{The difference in character between the layers is also evident from the dominant physical processes: in the TNTI turbulence propagates mostly via transmission of turbulent (i.e. Reynolds type) shear stresses~\citep{Westerweel2005, Westerweel2009}, whereas it is the action of viscous shear forces in Corrsin's theory.
\citet{Bisset2002} make an explicit distinction between the two layers by stating that the TNTI is a thin \emph{turbulent} layer connecting non-turbulent (irrotational) and the turbulent regions of the flow, and they conjecture that the VSL forms the outer boundary of the TNTI. The aims of this paper are firstly to determine whether a VSL can be observed for a generic shear flow and secondly to study in detail the structure of the turbulence boundary.

One important factor that may \rev{partly} explain the observed  differences in layer properties is the method by which the interface between the nonturbulent and turbulent fluid is identified.
Indeed, the interface is usually obtained by applying a threshold to a scalar field such as enstrophy~\citep{Bisset2002,Mathew2002,Holzner2007,DaSilva2008} or a high Schmidt number dye~\citep{Westerweel2005, Westerweel2009}.
By construction, this interface is artificial because the transition between turbulent and nonturbulent fluid must occur smoothly over a finite region.
The threshold value is generally chosen in a range where results are insensitive to the precise threshold value, e.g.\ for the conditional statistics \citep{Bisset2002, Holzner2007, Westerweel2005, Westerweel2009}.
A complication in this matter is} that in experiments and even in numerical data sets it is often difficult to vary thresholds over a span of several decades because of experimental \citep[e.g.][]{Westerweel2005,Holzner2007} or numerical \citep[e.g.][]{Bisset2002,Mathew2002} noise.

In this paper we perform a systematic study of the effect on the threshold value on the entrainment velocity and related statistics.
\rev{In doing so we span the entire range from essentially irrotational fluid near the turbulence boundary to fully turbulent fluid near the jet centre, which will enable us to address whether a VSL exists at the outer fringes of turbulence and how it may be related to a TNTI.}
Our analysis supports the existence of a VSL over a large range of thresholds ($\sim$ 20 decades), \rev{a turbulent core and a smooth transition zone connecting the two that will be identified as the buffer region. The buffer region shows several features characteristic of the TNTI}.
The study provides insight into how the local and global turbulent entrainment are connected. We find that the VSL is a nearly one-dimensional layer with low surface curvature and both its area and viscous transport velocity adjust to the imposed rate of entrainment so that the entrainment flux is independent of threshold.
We perform simulations of a temporal \rev{plane} jet at $\Re=5000$ which \rev{is a generic well-documented flow} \citep{DaSilva2008} \rev{that} is attractive from the viewpoint that \rev{it} has two homogeneous directions.
The present paper starts out with a theoretical framework (\S \ref{sec:theory}) that describes the properties of the local entrainment velocity from a local and a integral perspective. \rev{A simple conceptual model is developed that is } tested against the simulation data. After a brief explanation of the simulation setup (\S \ref{sec:simulations}), \rev{the entrainment characteristics, layer structure and geometry of the turbulence boundary} are presented (\S \ref{sec:results}). In \S \ref{sec:discussion}, the relation between the identified layer structure and the TNTI is explored, as well as the mechanism by which entrainment takes place. Concluding remarks are made in \S \ref{sec:conclusions}. In a companion paper the influence of mean shear and Reynolds number are analysed.

\section{Theory}
\label{sec:theory}
This section revisits \rev{the various definitions of entrainment velocity,} the determination of the local entrainment velocity based on enstrophy budgets from a local perspective and complements it with an integral approach to the problem. Thereafter a simplified model is set out with predictions for the enstrophy transport across the VSL.
\rev{
\subsection{Definitions of entrainment velocity}

One of the subtleties of turbulent entrainment is that there are several definitions for the entrainment velocity in use \citep{Turner1986, Hunt1983}. The most common definition is the rate at which fluid flows into the turbulent zone across its boundary, commonly denoted $E$. For the temporal jet, the specific flux $q(t)$ is constant and therefore $E=0$ (see also \S \ref{sec:simulations}).
A second definition is the rate at which the edge of a turbulent flow spreads outwards, i.e.\ the boundary entrainment rate $E_b$.
This is the quantity that is denoted in this work by $u_e$.
A third definition is $E_b^* = E_b - E$ which can be interpreted as the speed of the interface relative to the mean fluid velocity.
Note that $E$ and $E_b$ are global (macroscale) quantities measured in a fixed coordinate system \cite[sometimes called laboratory coordinates, e.g.][]{Westerweel2009}, whereas $E_b^*$ uses a coordinate system relative to the mean flow.
In the next section we will also define local (microscale) entrainment velocities $v_n$ that, similar to $E_b^*$, represent the local interface propagation velocity relative to the local fluid velocity.
}

\subsection{Local entrainment velocity: classical approach}
\label{sec:vL}
We \rev{differentiate} between turbulent and non-turbulent flow regions by using a threshold on the enstrophy $\omega^2 \equiv \omega_i \omega_i$, where $\omega_i$ is a component of vorticity.
This defines a bounding surface separating the two regions.
In a Lagrangian frame moving with the iso-enstrophy surface the isolevel will by definition remain constant and this property can be used to derive an expression for the propagation velocity~\citep{Holzner2011}.  We write the velocity of an isosurface element, $\boldsymbol{v}$, as a sum of fluid velocity, $\boldsymbol{u}$, and velocity of the area element relative to the fluid, $\boldsymbol{V}$, that is, $\boldsymbol{v}=\boldsymbol{u}+\boldsymbol{V}$.
The total change of $\omega^2$ in a frame of reference moving with \rev{an} enstrophy isosurface element is then given by
\begin{equation}
\frac{\d \omega^2}{\d t}=\frac{\partial\omega^2}{\partial
t}+\boldsymbol{v}\cdot\del\omega^2=\frac{D\omega^2}{D t}+\boldsymbol{V}\cdot\del\omega^2=0,
\end{equation}
where the lower case $\d / \d t$ is the total derivative following a surface element, and the upper case $D/Dt$ is the material derivative which follows a fluid element.
By defining a surface normal as $\boldsymbol{\hat{n}}=\del\omega^2/|\del\omega^2|$ and the normal relative velocity component $\hat{v}_n = \boldsymbol{V}\cdot\boldsymbol{\hat{n}}$, we obtain
\begin{equation}
\hat{v}_{n}=-\frac{1}{|\del\omega^2|} \frac{D\omega^2}{D t}.\label{eq:vnL0}
\end{equation}
Substituting the enstrophy balance equation
\begin{equation}
 \label{eq:enseq}
  \frac{D}{D t} \left( \frac{\omega^2}{2}\right) =
  \nu \del^2 \left( \frac{\omega^2}{2}\right)
  + \omega_i \omega_j s_{ij}
  - \nu \del \omega_i \cdot \del \omega_i
\end{equation}
into \eqref{eq:vnL0} and averaging over the isosurface $\langle \cdot \rangle_S$, we obtain an expression for the average entrainment velocity $v_n$:
\begin{equation}
\label{eq:vnL}
v_{n}\equiv\avS{\hat{v}_n}=v_L^{\mathcal{P}} + v_L^{\mathcal{D}} + v_L^{\mathcal{E}},
\end{equation}
where
\begin{equation*}
v_L^{\mathcal{P}} = -\avS{\frac{2\omega_i\omega_j s_{ij}}{|\del\omega^2|}}, \quad
v_L^{\mathcal{D}}=-\avS{\frac{\nu\del^2\omega^2}{|\del\omega^2|}}, \quad
v_L^{\mathcal{E}}=\avS{\frac{2\nu \del\omega_i \cdot \del\omega_i}{|\del\omega^2|}}.
\end{equation*}

\noindent Using the definition of $\vec{\hat{n}}$, the viscous term can be decomposed into a contribution due to curvature and normal transport through the following identity \citep{Holzner2011}:
\begin{equation}
  \label{eq:curvature}
  \del^2 \omega^2 = | \del \omega^2 | \del \cdot \vec{\hat{n}} + \vec{\hat{n}} \cdot\del | \del \omega^2 |.
\end{equation}
This identity will be used to quantify the role of curvature in \S \ref{sec:vn}.

\subsection{Local entrainment velocity: integral approach}
\label{sec:vn_integral}

An alternative to the local approach described in $\S$ \ref{sec:vL} is to integrate the enstrophy equation \eqref{eq:enseq} over a time-dependent domain $D(t)$ which has boundary velocity $\vec{v}$ and use the Reynolds transport theorem, resulting in
\begin{equation*}
\begin{split}
 \frac{\d}{\d t} \int_D \frac{\omega^2}{2} \d V + \oint_{\partial  D} (\vec{u} - \vec{v}) \cdot \vec{\hat{n}} \left( \frac{\omega^2}{2}\right)  \d S
  &= \nu \oint_{\partial  D} \del \left( \frac{\omega^2}{2}\right) \cdot \boldsymbol{\hat{n}} \d S \\
  & + \int_D \omega_i\omega_j s_{ij} \d V
  - \nu \int_D  \del\omega_i \cdot \del\omega_i \d V.
 \end{split}
\end{equation*}
\noindent As the surface normal $\boldsymbol{\hat{n}}$ points into the turbulent region, the appropriate volume under consideration comprises the irrotational region. We can formalise this by defining a control volume $D = H(1 - \omega^2/\omega_0^2)$ where $H$ is the Heaviside function and $\omega_0^2$ is an enstrophy threshold. As $\omega^2$ is then by definition constant on the surface $\partial D$, the equation above simplifies to
\begin{equation}
\begin{split}
\oint_{\partial  D} \vec{V} \cdot \vec{\hat{n}} \d S   =
 \frac{2}{\omega_0^2} & \left(  \frac{\d}{\d t} \int_D \frac{\omega^2}{2} \d V
- \nu \oint_{\partial  D} \del \left( \frac{\omega^2}{2}\right) \cdot \vec{\hat{n}} \d S \right. \\
  & \quad \left. - \int_D \omega_i\omega_j s_{ij} \d V
  + \nu \int_D  \del\omega_i \cdot \del\omega_i \d V \right).
 \end{split}
 \label{eq:vnL1}
\end{equation}
Using the Gauss divergence theorem and making use of the fact that $\oint \vec{u} \cdot \vec{\hat{n}} \d S = 0$ for an incompressible fluid, the entrainment flux $Q_e$ can be expressed as
\begin{equation}
\label{eq:vn}
Q_e = \frac{\d V}{\d t} = \oint_{\partial  D} \vec{V} \cdot \vec{\hat{n}} \d S \equiv v_n S, \quad\quad S = \oint_{\partial  D} \d S, \quad \quad V = \int_{D} \d V.
\end{equation}
\noindent Introducing an average over the volume $\avV{\cdot}$, \eqref{eq:vnL1} can be rewritten as
\begin{equation*}
\begin{split}
  \left( 1 - \frac{\avV{\omega^2}}{\omega_0^2} \right) v_n
  &= - \frac{2 \nu}{\omega_0^2 } \avS{\frac{\d \omega^2}{\d n}}
   + \frac{2 V}{S \omega_0^2} \left( \frac{\d}{\d t} \frac{\avV{\omega^2}}{2}
  - \avV{\omega_i\omega_j s_{ij}}
  + \nu \avV{\del\omega_i \cdot \del\omega_i} \right).
  \end{split}
\end{equation*}
\noindent Now, because $D$ spans the entire nonturbulent region, it is expected that $\avV{\omega^2}/\omega_0^2 \ll 1$, and therefore
\begin{equation}
  \label{eq:vnI}
  v_n  \approx v_{I}^{\mathcal{D}} + v_{I}^{\mathcal{T}} + v_{I}^{\mathcal{P}} + v_{I}^{\mathcal{E}},
\end{equation}
where
\begin{align*}
   v_{I}^{\mathcal{D}} &= - \frac{2 \nu}{\omega_0^2} \avS{\frac{\d \omega^2}{\d n}}, &
  v_{I}^{\mathcal{T}} &= \frac{2 V}{S \omega_0^2} \frac{\d}{\d t} \frac{\avV{\omega^2}}{2}, \\
  v_{I}^{\mathcal{P}} &= - \frac{2 V}{S \omega_0^2}
  \avV{\omega_i\omega_j s_{ij}}, &
  v_{I}^{\mathcal{E}} &= \frac{2 V}{S \omega_0^2} \nu \avV{\del\omega_i \cdot \del\omega_i}.
\end{align*}

\subsection{A model for enstrophy transport in the viscous superlayer}
\label{sec:vnprediction}
In the VSL, the evolution of enstrophy is governed by molecular processes  \citep{Corrsin1955, Holzner2011}, i.e. $|v^\mathcal{P} / v_n | \ll 1$. Assuming that the local curvature is small on average and multiplying by $\d \omega^2 / \d n$, Eq.~\eqref{eq:vnL} then becomes
\begin{equation}
 \frac{\d \omega^2}{\d n} v_n + 2 \nu \omega \frac{\d^2 \omega}{\d n^2} =
  2 \omega \frac{\d}{\d n} \left( v_n \omega + \nu \frac{\d \omega}{\d n} \right) = 0.
\end{equation}
\noindent Integrating this expression and using that at $n=-\infty$, both $\omega=0$ and $\d \omega / \d n = 0$, we obtain
\begin{equation}
  \label{eq:om2_ODE}
  v_n \omega + \nu \frac{\d \omega}{\d n} = 0.
\end{equation}
\noindent Assuming that $v_n$ is constant in the VSL, the square of the solution to \eqref{eq:om2_ODE} is
\begin{equation}
  \label{eq:om2_ansol}
  \frac{\omega^2}{\omega_r^2} = \exp\left(\frac{-2 v_n (n-n_r)}{\nu} \right),
\end{equation}
\noindent where $\omega_r^2 = \omega^2(n_r)$ is a reference value for enstrophy. Hence, the enstrophy is expected to drop off exponentially in the VSL, provided that $v_n$ is constant (see \S\ref{sec:vn}).

The model solution \eqref{eq:om2_ansol} can be used to predict the magnitude of the enstrophy transport terms. For the local approach we expect
\begin{align*}
  v_L^\mathcal{D}
  & \approx -\nu \frac{\d^2 \omega^2}{\d n^2} \left(\frac{\d \omega^2}{\d n}\right)^{-1} = 2 v_n, \\
   v_L^\mathcal{E} & \approx    2 \nu \left(\frac{\d \omega}{\d n}\right)^2 \left(\frac{\d \omega^2}{\d n}\right)^{-1}
    = - v_n.
\end{align*}
\noindent Entirely consistently, we expect for the integral approach that
\begin{align*}
   v_I^\mathcal{D} &\approx \frac{\nu}{\omega^2} \frac{\d \omega^2}{\d n} = 2 v_n, \\
  v_L^\mathcal{E} &\approx \frac{2 \nu}{\omega^2} \int_{-\infty}^{n} \left( \frac{\d \omega}{\d n} \right)^2 \d n = - v_n.
\end{align*}

\section{Simulations}
\label{sec:simulations}
The start situation for a temporal plane jet is a fluid layer that is quiescent except for a thin region $-b_0 < y < b_0$ where the streamwise velocity $u$ is nonzero, and is homogeneous in the two other directions $x$ and $z$. Here, $b_0$ is the initial jet width.
It follows from continuity that the volume flux $q = \int u \d y$ remains constant for this flow throughout the jet's transition to turbulence and subsequent growth due to turbulent entrainment.
Assuming that the Reynolds number $\Re \gg 1$, the only relevant parameters are the initial volume flux $q_0$ and time $t$ which suggests self-similar behaviour, with the jet halfwidth $b$ and centreline velocity $\hat{u}$ scaling as $b \propto \sqrt{q_0 t}$ and $\hat{u} \propto \sqrt{q_0 / t}$, respectively.

The simulation domain is a cuboid of size $24 b_0\times 36 b_0\times 24 b_0$, which is three times larger in all directions than the domain used in \cite{DaSilva2008}. The larger domain facilitates much longer simulations, thereby allowing not only the first moments but also the turbulence to reach an equilibrium and has the added advantage of improved statistics because of the larger area spanned by the two homogeneous directions.
\rev{Periodic boundary conditions are employed in the two homogeneous directions $x$ and $z$, and free-slip boundary conditions are imposed at $y=\pm 18 b_0$.}

The simulation considered here is for $\Rey \equiv 2 q_0 / \nu =5000$. The resolution of the simulation is $1024 \times 1536 \times 1024$ which is sufficient to ensure that all active length scales in the turbulence are fully resolved.
 We define a reference time-scale $t^* = b_0^2 / q_0$ and simulate for $300 t^*$. All statistics before $t/t^* = 150$ are discarded to ensure that the turbulence has time to reach a dynamic equilibrium.

Following \cite{DaSilva2002, DaSilva2008}, we use the initial condition
\begin{equation}
 u(y,0) = \frac{U_0}{2} \left[ 1 + \tanh\left(\frac{b_0-|y|}{2 \theta_0} \right) \right],
\end{equation}
\noindent where $U_0$ is chosen such that $\int u \d y = q_0$.
\rev{We set $\theta_0 = 2b_0/ 35$ \citep{DaSilva2008} and seed the initial condition with uniform random noise with an enstrophy level that is 8 percent of the maximum average enstrophy [which is $(U_0 / (4 \theta_0))^2$]. \rev{Note that the perturbation amplitude in terms of the velocity is only about one percent.} This facilitates a rapid transition to turbulence and we note that the enstrophy levels after the transition to turbulence far exceed the noise levels.} The code used for direct numerical simulation is based on fully conservative second order finite difference operators in space \citep{Verstappen2003} and uses an adaptive second-order Adams-Bashforth time integration scheme. \rev{The advantage of the spatial discretisation used is that the numerics are free of numerical diffusion whilst still satisfying volume and momentum conservation}. More details can be found in \cite{vanReeuwijk2008a}.

As the statistics shown in the next section depend heavily on budgets for the enstrophy, special care is taken to ensure that the budgets are calculated consistently with the numerics used.
To achieve this, a mimetic \citep{Hyman1997, vanReeuwijk2011} \texttt{curl} operator is defined such that it satisfies the identity $\nabla \times \nabla p = 0$ up to machine precision, where $p$ denotes pressure. In order to ensure calculation fully compatible with the numerical method, we do not manually discretise \eqref{eq:enseq}, but instead make use of the following identities
\begin{gather}
  \label{eq:I1}
  \omega_i \omega_j s_{ij} = (\vec{u} \cdot \del) \frac{\omega^2}{2} - \vec{\omega} \cdot (\del \times (\vec{u} \cdot \del) \vec{u}), \\
  \label{eq:I2}
\nu \del \omega_i \cdot \del \omega_i = \nu \del^2 \left( \frac{\omega^2}{2}\right) - \nu \vec{\omega} \cdot (\del \times \del^2 \vec{u}),
\end{gather}
\noindent which are then also enforced on the discrete level. Taking \eqref{eq:I2} as an example, one can calculate the first term on the right hand side directly by using the routine for scalar diffusion on $\omega^2/2$; the second term can be calculated by taking the discrete curl of the viscous term in the momentum equation, and then taking the scalar product of the result with the vorticity components.

As the temporal jet has a nonzero mean velocity in the $x$-direction, it is important to ensure that the identity $\oint \vec{u} \cdot \vec{\hat{n}} \d S = 0$ is also satisfied on the discrete level.
Indeed, this identity was used explicitly to derive \eqref{eq:vn}.
This can only be achieved if the thresholding algorithm identifies entire cells to be either inside or outside the turbulent region.
Indeed, we found that if we used trilinear interpolation to construct an isosurface - which is in principle a better representation - the various interpolations required could lead to very significant deviations in the calculated entrainment velocity.

\section{Results}
\label{sec:results}

\subsection{Bulk flow properties}

\begin{figure}
\centering
\includegraphics{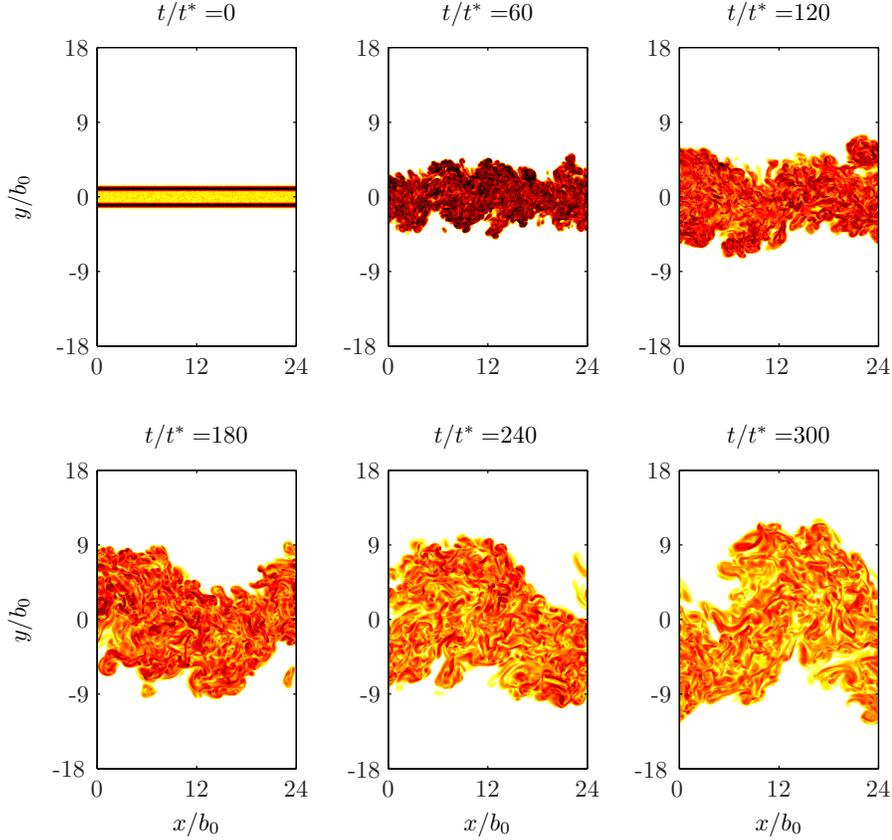}
\caption{\label{fig:evolution}: Jet development as indicated by $^{10}\log(\omega^2/\omega_r^2)$. The color range is from $\omega^2 / \omega_r^2 = 10^{-4}$ (white) to $\omega^2 / \omega_r^2 = 10^{1}$ (black). A colorbar is shown in Fig.\ \ref{fig:thresholds}.}
\end{figure}

\rev{The time evolution of the enstrophy levels $^{10}\log (\omega^2/\omega_r^2)$ in the jet are shown in Fig.\ \ref{fig:evolution}. Here, $\omega_r^2 \equiv U_0^2 / b_0^2$ is a reference enstrophy level. Fig.\ \ref{fig:evolution}(top left) shows the initial condition and the low-amplitude noise. At $t/t^* = 60$, the flow has fully transitioned to turbulence and high enstrophy levels can be observed in the jet that very rapidly drop off near the jet edge. As time progresses further, the enstrophy levels decrease and spatial scales can be seen to grow.}

\begin{figure}
 \centering
 \includegraphics{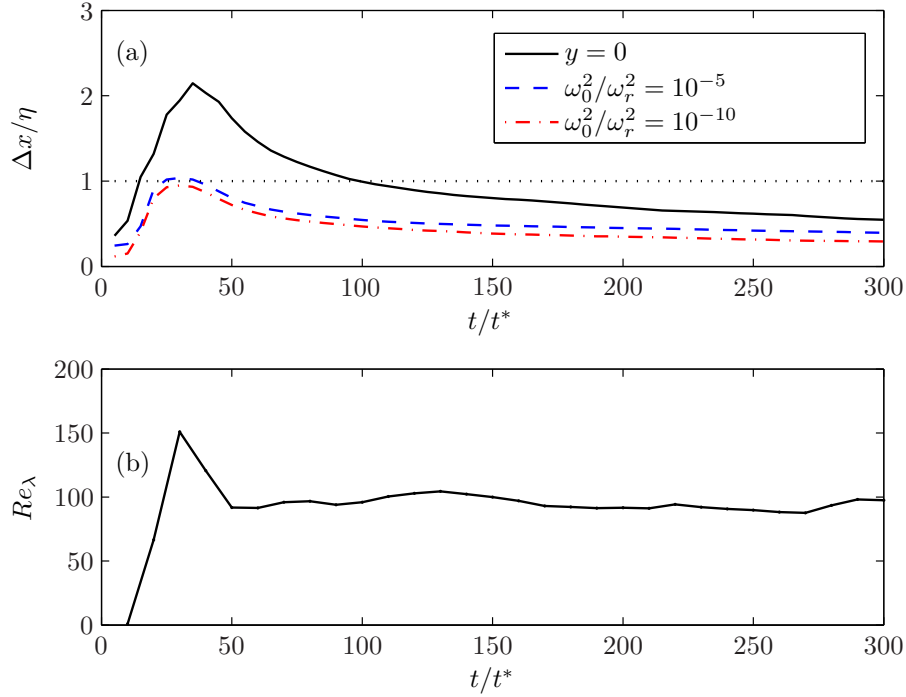}
 \caption{\label{fig:quality} (a) $\Delta x$ normalised by the Kolmogorov scale $\eta$;  (b) $\Rey_\lambda$ against time.}
\end{figure}

\begin{figure}
 \centering
 \includegraphics{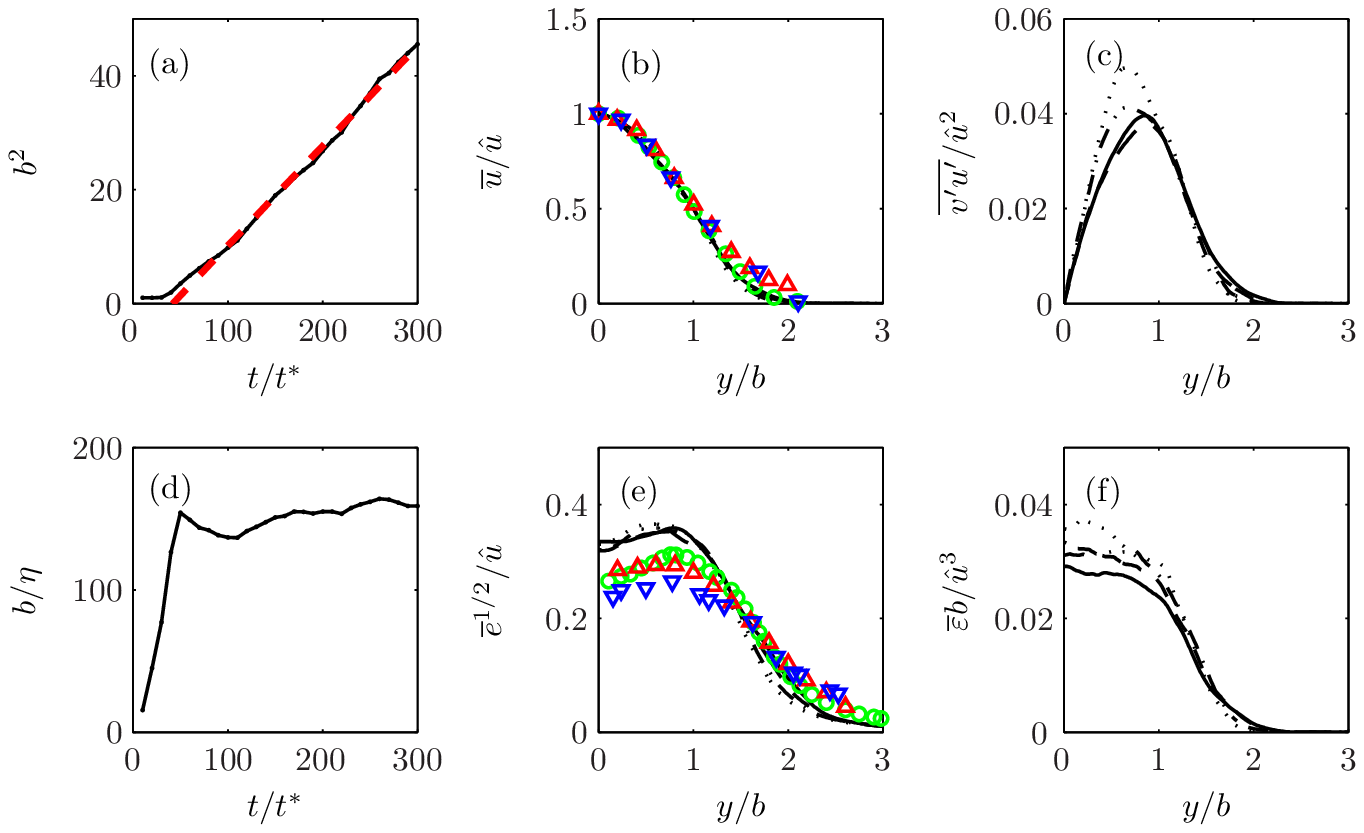}
 \caption{\label{fig:SSjet} Self-similarity of the temporal plane jet. (a) dependence of $b^2$ on time; (b) average velocity $\av{u}$; (c) turbulent momentum flux $\av{\fl{u}\fl{v}}$; (d) dependence of $b/\eta$ on time. (e) kinetic energy $\overline{e}$; and (f) the dissipation rate $\overline{\varepsilon}$. Also shown in (b) and (e) are the data from \cite{DaSilva2008} (green circles), \cite{Gutmark1976} (red upward triangles) and \cite{Ramparian1985} (blue downward triangles). \rev{The time-sequence in (b,c,e,f) are averages over the interval $t/t^*=$ 150-190 (---), 190-230 ($-\ -$), 230-270 ($-\cdot-$) and 270-300 ($\cdots$), respectively.}}
\end{figure}

Fig.~\ref{fig:quality}(a) demonstrates that the grid resolution is appropriate for the problem under consideration. Shown is the grid spacing normalised by the Kolmogorov length scale $\eta=(\nu^3/\varepsilon)^{1/4}$ based on the centreline dissipation rate $\varepsilon(t) \equiv \overline{\varepsilon}(y=0, t)$.
The overbar denotes averaging over the two homogeneous directions and over 10 $t^*$.
The dissipation rate has its maximum at $y=0$ and dissipation rates will be much lower at the jet boundary, which implies that the simulation is even better resolved there (dashed and dash-dotted lines).
As can be seen, the simulation becomes better resolved in time because $\eta \propto \sqrt{t}$, which can be inferred by using $\varepsilon \propto \hat{u}^3/b$, as is confirmed in Fig.~\ref{fig:SSjet}(f).

Fig.~\ref{fig:quality}(b) shows the evolution of the Taylor Reynolds number, \rev{defined as $\Rey_\lambda = (2 e / 3)^{1/2} \lambda / \nu$, where $e(t) \equiv \overline{e}(y=0, t)$ is the centreline turbulent kinetic energy and $\lambda = \sqrt{10 \nu e / \varepsilon}$ is the Taylor microscale \citep[e.g.][pp. 67-68]{Tennekes1972}.}
As judged from $\Rey_\lambda$, the turbulence reaches an equilibrium value of about $\Rey_\lambda \approx 100$ after $t/t^*=50$.

\begin{figure}
 \centering
 \includegraphics{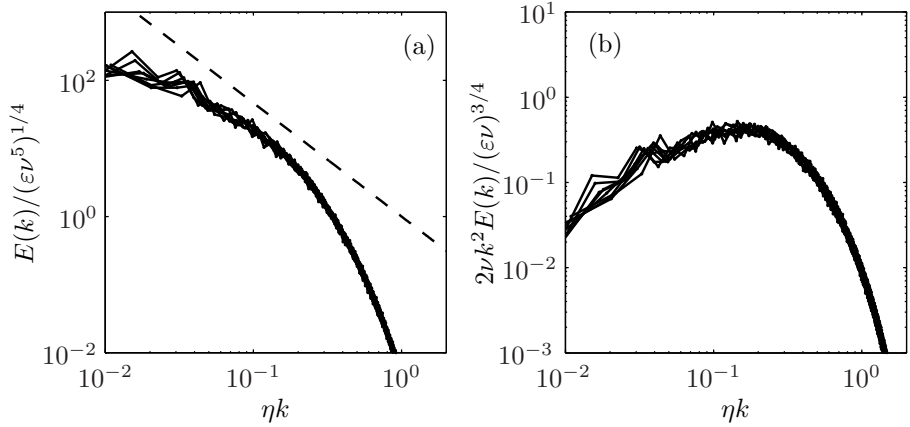}
 \caption{\label{fig:spectra}Energy density (left) and dissipation spectra (right) at $y=0$.}
\end{figure}

Energy density spectra for the plane $y=0$, averaged over shells of wave number $\sqrt{k_x^2+k_z^2}$ and a time interval of $10 t^*$ are shown in Figure~\ref{fig:spectra}.
Fifteen spectra are shown for $t/t^* > 150$ and the collapse demonstrates the self-similarity of the flow under consideration; even though the spectra change in time, the normalisation with $\eta$ and $\varepsilon$ cause a full collapse of the data.
The spectra indicate that there is a range of active scales and that there is a \rev{small} separation of scales as is evident from the formation of a $k^{-5/3}$ spectrum (left) and a \rev{comparison of the peak in the energy density (left) and} dissipation spectrum (right).
\rev{Note that the dissipation spectrum peaks at $k \eta \approx 0.2$, once more indicating that the simulation is fully resolved.}

As the velocity profile is symmetric around $y=0$, the jet half-width $b$ was inferred from the average of the values of $b$ for which $\av{u}(b,t)=\hat{u}(t)/2$ and $\av{u}(-b,t)=\hat{u}(t)/2$.
For all profiles shown, use has been made of the symmetry (or anti-symmetry) in the profile to further improve the statistical accuracy. Shown in Fig.~\ref{fig:SSjet}(a) is the scaling of $b^2$ with time, which as expected from dimensional arguments is linear; the red dashed line is a linear fit.
The normalised mean velocity $\av{u}$ and momentum flux $\av{\fl{u}\fl{v}}$ are shown in Fig \ref{fig:SSjet}(b,c). These profiles were scaled and then further averaged over four contiguous time-intervals (an effective average over 40 $t^*$). The profiles are convincingly self-similar\rev{, although the profile of $\av{\fl{u}\fl{v}}$ shows more variability than $\av{u}$ because it is a second order moment.
Also shown in Fig \ref{fig:SSjet}(b) is data from laboratory experiments of a plane jet by \cite{Gutmark1976} (red upward triangles) and \cite{Ramparian1985} (blue downward triangles) as well as the numerical simulations of a temporal plane jet by \cite{DaSilva2008} (green circles).
Good agreement can be observed.
}
Self-similarity of turbulent quantities is demonstrated in Figs \ref{fig:SSjet}(d-f).
As discussed earlier, the balance between turbulence production and dissipation suggests that $\overline{\varepsilon} \propto \hat{u}^3/b$ (Fig.~\ref{fig:SSjet}(f)), which in turn implies that $\eta \propto b$ (Fig.~\ref{fig:SSjet}(d)).
The profiles for the turbulence kinetic energy (TKE) $\overline{e}$ and the dissipation rate $\overline{\varepsilon}$ (Figs \ref{fig:SSjet}(e, f)) show once more a reasonably good collapse, with $\overline{e}$ peaking at $y/b \approx 1$ (where shear-production is maximal) and the dissipation rate $\overline{\varepsilon}$ peaking at the centreline.
We observed variations in the growth rate of $b$ between different simulations \rev{upon variation of the initial conditions}, despite a convincing self-similar behaviour in all of them. This may point to non-universal self-similar behaviour.
\rev{Indeed, \cite{Redford2012} showed through simulations of an axisymmetric temporal wake that differences in the initial conditions can influence growth rates (and therefore entrainment coefficients) for extended periods.  During such  periods, the flow developed in a self-similar fashion but was not universal; only on very large timescales was universal behaviour observed.
The non-universal self-similarity may explain the  slightly higher turbulence levels in the current simulation compared to those observed by others (Fig. \ref{fig:SSjet}(e)).}

\subsection{Entrainment velocity and budgets}
\label{sec:vn}
\begin{figure}
 \centering
 \includegraphics{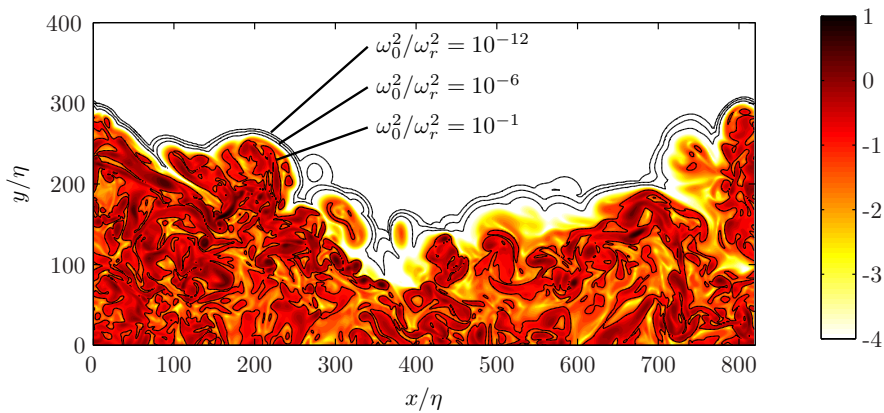}
 \caption{\label{fig:thresholds} A field in cross-section showing $^{10}\log (\omega^2/\omega_r^2)$ at $t/t^*=150$. Also shown are various isocontours of enstrophy.}
\end{figure}

\rev{A cross-section of the enstrophy field at constant $z$ is shown in Fig.~\ref{fig:thresholds}, together with  enstrophy isolines at $\omega_0^2/\omega_r^2=10^{-12}, 10^{-6}$ and $10^{-1}$.
At $\omega_0^2/\omega_r^2=10^{-1}$, the turbulent/nonturbulent interface is highly contorted and has 'holes', whereas the lower thresholds do not have such holes.
Note that what may appear as holes on the figure~(i.e. in 2D) for low tresholds are fluid portions that are connected in 3D to the outer irrotational region, while at higher thresholds one also finds islands of low vorticity disconnected from the outer region.
What is striking is that the enstrophy levels at  $\omega_0^2/\omega_r^2=10^{-12}$ and $10^{-6}$ are in close proximity to the high enstrophy regions, indicating a very quick drop-off of enstrophy levels near the \rev{turbulence boundary}. At these low threshold values, the surface remains contorted because of the large scale vortices distorting the flow but the enstrophy isosurfaces appear to form nearly one-dimensional layers with relatively small curvature.}

The instantaneous budgets of enstrophy were calculated for the entire 3-D field every 5 $t^*$ and then used to calculate the terms in \eqref{eq:vnL}, \eqref{eq:vnI} for 37 thresholds in the range $\omega_0^2/\omega_r^2 \in [10^{-24}, 10^0]$.
Simultaneously, the volume $V$ was recorded for each of the threshold values, which enabled an independent calculation of the propagation velocity $v_n$ using \eqref{eq:vn}.
Shown is the data for $t/t^* > 150$.

\begin{figure}
 \centering
 \includegraphics{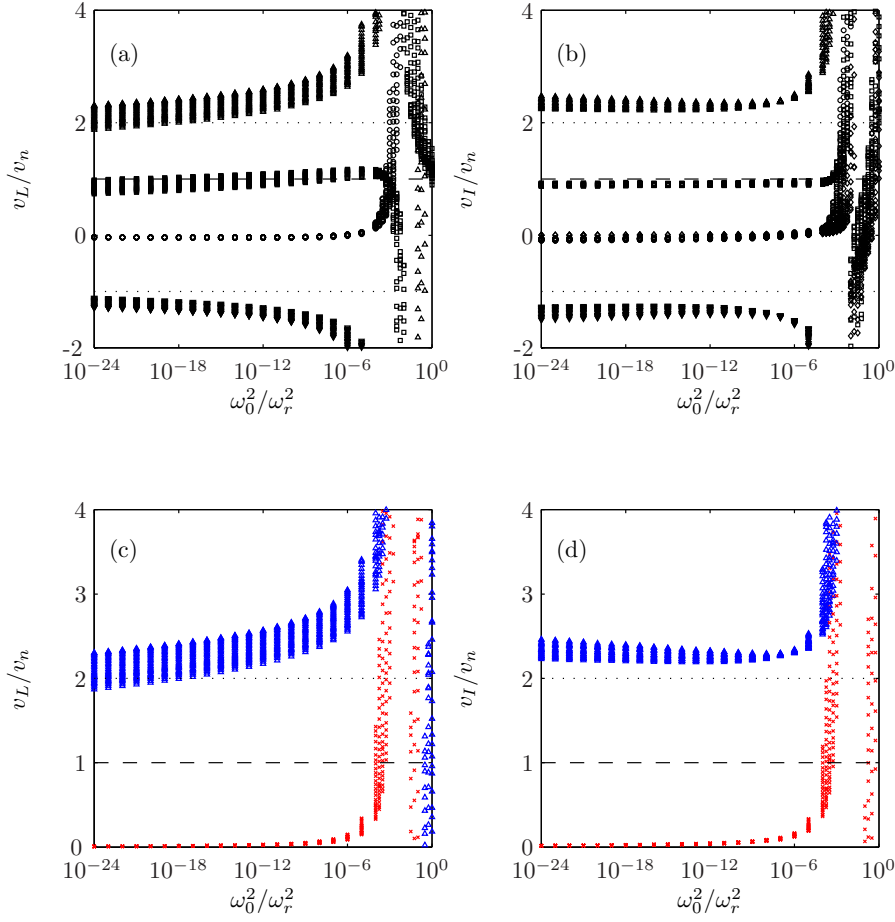}
 \caption{\label{fig:vndec_vn} the local entrainment velocity decomposed into various contributions and normalised by the actual propagation velocity $v_n$. $v^\mathcal{D}$ [\CaptionUpTriOpen], $v^\mathcal{P}$ [\CaptionCircleOpenSml],   $v^\mathcal{E} $ [\CaptionDownTriOpen], $v^\mathcal{T}$ [\CaptionDiamondOpen], total [\CaptionSquareOpen]. (a,c) $v_L$; (b,d) $v_I$. Figs (c,d) show the decomposition of $v^\mathcal{D}$ in a contribution due to curvature (red crosses) and normal transport (blue triangles).}
\end{figure}

\rev{First, we show that the calculated local entrainment velocities $v_L$ and $v_I$ correspond to the actual local entrainment velocity $v_n$.
To this end, the terms comprising the local entrainment velocity $v_L$ \eqref{eq:vnL} and $v_I$ \eqref{eq:vnI} normalised by the directly measured volume-based entrainment velocity $v_n$ are presented in Fig.~\ref{fig:vndec_vn}. For $\omega_0^2/ \omega_r^2<10^{-5}$, the predicted propagation velocities $v_L$ (squares, Fig.~\ref{fig:vndec_vn}(a)) and $v_I$ (squares, Fig.~\ref{fig:vndec_vn}(b)) match the actual propagation velocity excellently.
A small systematic error can be discerned in the calculation of $v_L$, as the calculated propagation velocity shows a small but systematic trend in $\omega_0^2$.
This systematic trend is not observed in $v_I$, although the prediction is slighty lower than $v_n$.
The poor predictions for $\omega_0^2/\omega_r^2>10^{-5}$ are not associated with $v_L$ and $v_I$ but are due to an insufficient temporal sampling frequency creating large errors in the calculation of $v_n \propto \delta V / \delta t$ (Eq. \ref{eq:vn}); this occurs because the high enstrophy regions tend to shrink and expand rapidly on a timescale shorter than $5 t^*$.
}

\rev{Fig.~\ref{fig:vndec_vn}(a) shows convincing evidence of the existence of a viscous superlayer (VSL). Indeed, in the VSL the inviscid contribution $v_L^\mathcal{P}$ (circles) does not play a role \citep{Holzner2011} and we observe that this is the case for $\omega_0^2/\omega_r^2 < 10^{-5}$.}
For $\omega_0^2/\omega_r^2 > 10^{-5}$, the inviscid terms increase very rapidly.
Also shown in Fig.~\ref{fig:vndec_vn}(a) are the theoretical predictions from \S\ref{sec:vnprediction}, namely $v_L^\mathcal{D} = 2 v_n$ and $v_L^\mathcal{E} = -v_n$ (both displayed by dotted lines).
The predictions are in good agreement, although the observed magnitudes of $v_n^\mathcal{E}$ (downward triangles) and $v_n^\mathcal{D}$ (upward triangles) are a bit larger \rev{than predicted, and a dependence on $\omega_0^2$ is discernable}.

\rev{The budget of $v_I$ is similar to that of $v_L$:} the temporal and inviscid contributions, $v_I^\mathcal{T}$ (diamonds) and $v_I^\mathcal{P}$ (circles) respectively, are negligible and the propagation of the enstrophy isosurface in the VSL is caused by viscous effects. For this indicator, the contributions $v_I^\mathcal{D}$ and $v_I^\mathcal{E}$ seem to become fully independent of the threshold level below $\omega_0^2/\omega_r^2 < 10^{-5}$.

One of the main assumptions made in the derivation of the theoretical model was that the curvature of the isosurface is small. The reasonably good agreement with the theoretical model in Fig.~\ref{fig:thresholds} supports this assumption but by using Eq.\ \ref{eq:curvature} it can be validated explicitly.
Figs \ref{fig:vndec_vn}(c,d) show the contribution to $v^\mathcal{D}$ by curvature (red crosses) and diffusive transport in the direction of the surface normal (blue triangles). For both $v_L$ as $v_I$, the curvature term becomes negligible for $\omega_0^2/\omega_r^2 < 10^{-5}$.
This provides confirmation that the theoretical model in \S \ref{sec:vnprediction} is a reasonable description of the processes governing the VSL.

\begin{figure}
 \centering
 \includegraphics{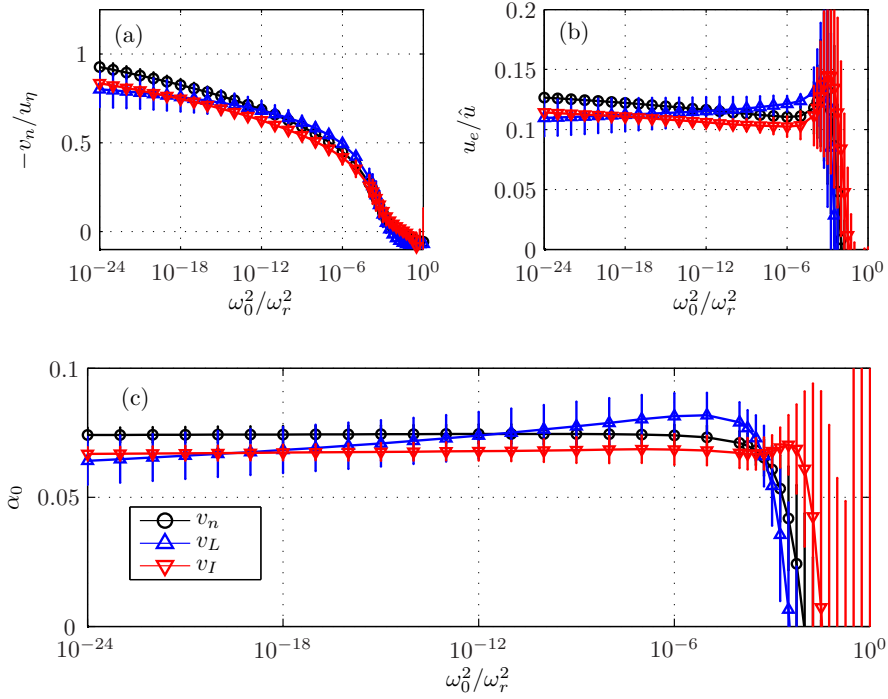}
 \caption{\label{fig:vn} Entrainment velocities as a function of threshold. (a)  $v_n$ normalised by $u_\eta$. (b) $u_e$ normalised by $\hat{u}$. (c) normalised entrainment coefficient $\alpha_0$.}
\end{figure}

Having established that both estimates of the local entrainment velocity $v_L$ and $v_I$ are in good agreement with the actual entrainment velocity $v_n$, we study the dependence of the entrainment velocity on the threshold value $\omega_0^2$.
In Fig.~\ref{fig:vn}(a) we show the dependence of $v_n$, $v_L$ and $v_I$ on $\omega_0^2$\rev{, normalised by the Kolmogorov velocity scale $u_\eta$.
The vertical error bars denote the variation over the entire time interval $150 < t/t^* < 300$.}
There is a clear dependence of $v_{n}$ on the enstrophy threshold: isosurfaces for the very low enstrophy thresholds propagate faster than those with higher thresholds.
Indeed, $v_{L}$ is nearly twice as high  at $\omega_0^2/\omega_r^2 = 10^{-24}$ than at $\omega_0^2/\omega_r^2 = 10^{-6}$.
Hence, although $v_n$ is of the same order of magnitude as $u_\eta$, the dependence on $\omega_0^2$ suggests that it is not merely the viscosity $\nu$ and dissipation rate $\varepsilon$ that determine the propagation velocity in the VSL.
\rev{This may be a low-Reynolds number effect: as the Reynolds number increases, the VSL will become thinner relative to the integral scale $b$ and therefore the enstrophy levels in the VSL will drop off quicker, cf.~\eqref{eq:om2_ansol}, leaving less opportunity for variation in $S$.
It should also be noted that the surface $S$ is not smooth but follows the grid (cf. \S \ref{sec:simulations}) to ensure conservation properties.
Further work is required to settle this issue.}

Another striking feature is that $v_n$ becomes zero around $\omega_0^2 / \omega_r^2 \approx 10^{-3}$ and \rev{positive} for $\omega_0^2 / \omega_r^2 > 10^{-3}$.
Hence, high enstrophy regions move inwards towards the jet centre, low enstrophy regions move outwards\rev{ and there exists an isosurface that separates the shrinking and expanding regions.
The movement of high enstrophy regions towards the jet centre} can be explained by using the relation between enstrophy and the dissipation rate $\varepsilon = \nu \av{\fl{\omega}\fl{\omega}}$ which is valid for isotropic and homogeneous turbulence .
Using the self-similarity of $\varepsilon$ it follows that $\av{\fl{\omega}\fl{\omega}} \approx \hat{u}^3 / (\nu b) \propto t^{-2}$. Hence, if one would assume that $\av{\fl{\omega}\fl{\omega}}$ has a self-similar profile and pick a reference threshold $\omega_0^2$ that one follows in time, it would be seen to move inwards towards the jet core. This applies to enstrophy levels where turbulence is developed and $v_n$ is positive. Towards the VSL, at low $\omega^2$ levels, viscous transport is diffusing enstrophy outwards and $v_n$ is negative. Note that the gradient of $-v_n$ over $\omega^2$ in Fig.~\ref{fig:vn}(a) is always negative meaning that the enstrophy profile is flattening over time.

\rev{
As macroscale entrainment is independent of molecular processes, it is expected that the
global entrainment velocity $u_e$ is independent of threshold.
As mentioned in \S \ref{sec:theory}, for the case under consideration $u_e$ corresponds to the boundary velocity $E_b$ in the terminology of \cite{Turner1986} and is defined as $u_e = L^{-2} \d V / \d t$ where $L=24 b_0$.
The relation to $v_n$ (Eq.~\ref{eq:vn}) is therefore $u_e = -(S/L^2) v_n$.}
Fig.~\ref{fig:vn}(b) shows the global entrainment velocity $u_{e}$ normalised by the centreline velocity amplitude $\hat{u}$.
The standard entrainment assumption \citep[e.g.][]{Turner1986} is $u_e = \alpha \hat{u}$ and the ratio plotted is the entrainment coefficient $\alpha$.
As can be seen, $\alpha$ is \rev{practically constant as a function of  $\omega_0^2$ although there is a small systematic trend.
When $\alpha$ is independent of $\omega_0^2$,} it implies that the entrainment \emph{flux} $Q_e$ is constant in the VSL; indeed, by using $u_e = v_n S / L^2 = Q_e / L^2$ it immediately follows that $Q_e$ is constant if $u_e$ is independent of $\omega_0^2$.
This result indicates that to first order $S \propto v_n^{-1}$, and only if $v_n \propto u_\eta$ is independent of $\omega_0^2$ do we expect that $S \propto \Rey^{1/4}$ independently of $\omega_0^2$, i.e. the classical view advocated by Corrsin.

\rev{The small trend discernable in Fig.~\ref{fig:vn}(b) is a low-Reynolds effect associated with the position of the interface.
Indeed,} selfsimilarity implies that $\av{u} = \hat{u} f(\xi)$, where $f$ is the universal velocity profile and $\xi = y / b$ the selfsimilarity variable.
The entrainment assumption is $u_{e0} = \alpha_0 \hat{u}$, where $u_{e0}$ is defined as $u_{e0} \equiv \d b / \d t$ and in our case $b$ is the half-velocity width.
As mentioned in \S\ref{sec:introduction}, other definitions of $b$ will result in different values for $\alpha$.
This is straightforward to see by fixing $\xi$ at a value differing from unity. Indeed, for an alternative width $y=\xi b$, we obtain
\begin{equation}
 u_e = \frac{\d y}{\d t} = \xi \frac{\d b}{\d t} = \alpha_0 \xi \hat{u},
\end{equation}
indicating that the effective entrainment rate is $u_e /\hat{u} = \xi \alpha_0$.
Hence, if there is a dependence of the average interface position $\avS{y}$ on $\omega_0^2$ this will create a trend in $u_e$.
In Fig.~\ref{fig:vn}(c), we have plotted $\alpha_0 = u_e / (\xi \hat{u})$, where $\xi = \avS{y} / b$ and $\avS{y}$ is the mean $y$-position of the isosurface.
\rev{As can be seen, the value of $\alpha_0$ is constant for both the directly measured entraiment velocity $v_n$ and the calculated entrainment velocity $v_{I}$.
There is a small downward trend for the value of $\alpha_0$ as calculated from $v_{L}$ which is due to the small systematic error discussed earlier.
The dependence of $\avS{y}$ on $\omega_0^2$ will no longer play a role at very high Reynolds numbers because the VSL will become so thin that $\avS{y}$ will become independent of $\omega_0^2$.}

\rev{In summary, the dependence of the variation of $v_n$ on $\omega_0^2$ can be explained by two independent mechanisms: 1) the dependence of the surface area $S$ on $\omega_0^2$; and 2) the finite thickness of the turbulence boundary as scaled on the jet thickness $b$.
These are likely to be low-Reynolds-number effects and it is expected that in the limit of $\Rey \rightarrow \infty$, the classical Corrsin viewpoint will be recovered.
Further work is required to verify this hypothesis.
}

\begin{figure}
 \centering
 \includegraphics[width=120mm]{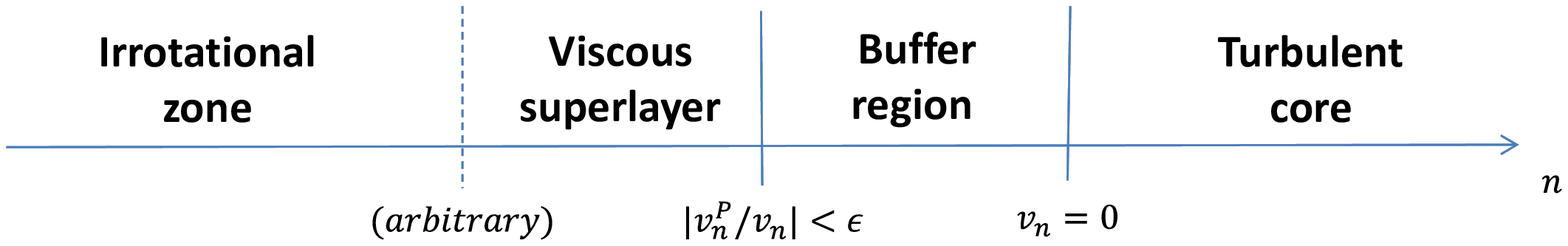}
 \caption{\label{fig:layerstructure} Definition sketch of layer structure of the turbulence boundary.}
\end{figure}

\rev{
\subsection{The structure of the turbulence boundary}

Having access to the entrainment velocity as a function of the threshold value $\omega_0^2$ allows one to explore the structure of the turbulence boundary.}
There are two distinguishing enstrophy threshold values to consider at any time, namely (i) the enstrophy level at which $v_n = 0$ and (ii) the enstrophy level for which enstrophy production becomes negligible,  diagnosed by the  criterion \rev{ $ | v_n^\mathcal{P} / v_n | < \epsilon$, where $\epsilon$ is a small value.
\rev{As discussed in the previous section, the} level $v_n = 0$ is a threshold that separates expanding regions ($v_n < 0)$ from shrinking regions ($v_n > 0$) and the threshold $ | v_n^\mathcal{P} / v_n | < \epsilon$ demarcates the start of the VSL.
This suggests a layer structure as shown in Fig.\ \ref{fig:layerstructure}. Away from the jet boundary, the flow is non-turbulent and irrotational. Moving closer to the turbulence boundary, one enters the VSL. The transition location is arbitrary and would depend on a choice of threshold.
The VSL extends up to the location where inviscid terms start playing a role.
Note that, similar to the viscous sublayer in a wall-bounded flow, the VSL \rev{can be classified neither as turbulent nor as laminar} because viscous effects are dominant whilst there are significant fluctuations in the layer due to external influences.
We define the turbulent core (TC) to be the region for which $v_n>0$, and  define a buffer region (BR) to be the zone between the VSL and the TC, which is still expanding but for which inviscid terms are important.
This term was chosen to emphasise the resemblance with the buffer layer of a wall-bounded turbulent flow that also couples two regions in which different processes dominate (viscous and inertial effects in the case of a wall-bounded flow).
}

\begin{figure}
 \centering
 \includegraphics{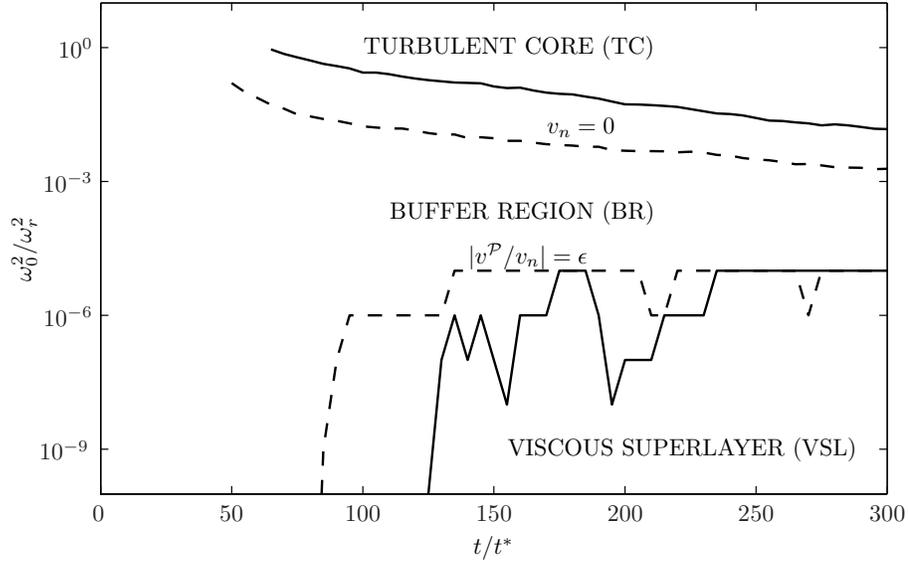}
 \caption{\label{fig:regions} A map of the different regions as identified by the thresholds on enstrophy; $v_L$ (dashed lines) and $v_I$ (solid lines).}
\end{figure}

\rev{The evolution of the layer structure} is plotted as a function of time in Fig.~\ref{fig:regions} for $v_{L}$ (dashed lines) and $v_{I}$ (solid lines).
\rev{A value  $\epsilon= 0.05$ was adopted to demarcate the onset of the VSL.
}
Although the exact values obtained by the two estimates differ, the trends are consistent.
The three regions are clearly visible in Fig.~\ref{fig:regions}.
As time progresses, the threshold level for which $v_n = 0$ moves to lower and lower \rev{values}, in accordance with the decay expected from self-similarity.
The threshold level demarcating the beginning of the VSL remains approximately constant at $\omega_0^2/\omega_r^2 \approx 10^{-5}$ for $t/t^* > 150$, \rev{although significant fluctuations can be observed for $t/t^*<250$ and the threshold values differ by a factor $10$ in the range where the VSL resides. Hence, it is impossible to infer unambiguously whether the onset of the VSL occurs at a fixed threshold value or not.
}

\rev{
\subsection{Geometry: relating $\omega_0^2$ to a distance}
}
The dependence of volume $V$ and surface area $S$ on the threshold value $\omega_0^2$ can be used to obtain information about the average distance from one enstrophy isosurface to the next, thereby getting an impression of the distance between different regions of the flow.
Making use of the observation that curvature is low at low thresholds, we can define a average distance $n$, which is related to the volume $V_T$ for which $\omega^2 > \omega_0^2$ and the surface area $S$ as
\begin{equation}
  \frac{\d V_T}{\d n} = S.
\end{equation}
Note that the sum of $V$ and $V_T$ is exactly half of the simulation domain volume.
Introducing $\Delta V_T = V_{T;i+1} - V_{T;i}$ and $\Delta n = n_{i+1} - n_{i}$ as the difference in volume and distance respectively between two adjacent enstrophy thresholds $\omega_{0;i}^2$ and $\omega_{0;i+1}^2$, the average distance can be approximated by $\Delta n \approx 2 \Delta V_T / (S_i + S_{i+1})$.
We set $n=0$ at the start of the VSL at $\omega_0^2/\omega_r^2 = 10^{-5}$.

\begin{figure}
 \centering
 \includegraphics{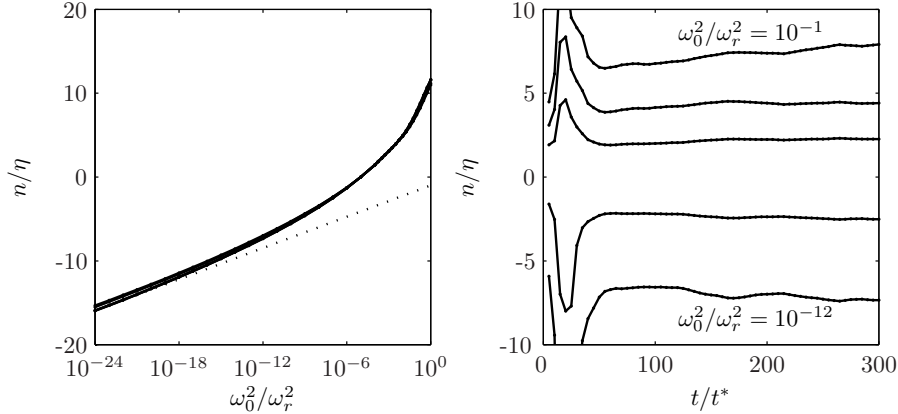}
 \caption{\label{fig:VSLthickness} (a) thickness as a function of threshold value; (b) thickness as a function of time.}
\end{figure}

The dependence of $n$ on $\omega_0^2$ is shown for $t/t^*>150$ in Fig.~\ref{fig:VSLthickness}(a). The collapse of the profiles for different times shows that the result is quite robust.
It is striking how close essentially irrotational regions are located to regions with very high enstrophy levels on average.
\rev{Indeed, the BR, which connects the TC and the VSL, is on average about $10 \eta$ or $0.5 \lambda$ thick.}
Moreover, the full 24 decades in enstrophy levels are separated by $25 \eta$ or $1.25 \lambda$.
Also shown is the analytical solution \eqref{eq:om2_ansol} (dotted line), expressed as $\omega_0^2 / \omega_r^2 = \exp( \alpha_\eta (n-n_r) / \eta)$ with $\alpha_\eta = 3.7$.
It is evident that \eqref{eq:om2_ansol} is a reasonable approximation for $\omega_0^2 / \omega_r^2 \ll 1$, which explains the agreement of the simulation data with the predictions for $v^\mathcal{D}$ and $v^\mathcal{E}$.
However, as can be seen, the shape of the $n$ profile deviates from a straight line in a semilog plot because of the influence of surface area $S$ on $v_n$.
Indeed, the model assumes that $v_n$ is constant and this has been clearly shown not to be the case for this flow at the given Reynolds number $\Re=$5000.
Fig.~\ref{fig:VSLthickness}(b) shows the normalized distance for a number of thresholds as a function time.
After $t/t^* = 150$, the distances become nearly constant, indicating that the jet edge has reached a dynamic equilibrium and the distances between isosurfaces scale with $\eta$.

\rev{
\section{Discussion}
\label{sec:discussion}

\subsection{Relation with the TNTI}
In the previous section we explored the properties of the turbulence boundary using a large range of enstrophy thresholds and identified a viscous superlayer, a buffer region and a turbulent core. In this section, the relation to the TNTI is established.

In their studies of the TNTI, \cite{Bisset2002,DaSilva2008, Mathew2002} used an enstrophy-based threshold of $\omega_0^2/\omega_r^2=0.1$. Based on Fig.\ \ref{fig:regions}, this would correspond to an interface located roughly on the boundary between the buffer region and the turbulent core.
However, a direct comparison of the value of the thresholds might not be the best way to ascertain where the TNTI resides.
Indeed, the enstrophy levels are both time-dependent and Reynolds-number dependent.
This can be made explicit by using the relation $\varepsilon \sim \nu \av{\fl{\omega}\fl{\omega}}$ valid for homogeneous turbulence, using $\varepsilon\sim \hat{u}^3/b$  (Fig.\ \ref{fig:SSjet}) and substituting the definition for $\Rey$, which yields
\begin{equation}
  \frac{\av{\fl{\omega}\fl{\omega}}}{\omega_r^2} \sim \frac{\varepsilon}{\varepsilon_0} \Rey.
\end{equation}
Since $\varepsilon \sim t^{-2}$ for this particular flow , the relation above implies that $\av{\fl{\omega}\fl{\omega}} \sim \Rey ~t^{-2}$ and enstrophy levels thence depend both on time and on the Reynolds number.
As the simulations performed in the present work are much longer than is usual, the enstrophy levels will be different than those reported by others.
Indeed, if one compares the isosurface of $\omega_0^2/\omega_r^2=0.1$ in Fig. \ref{fig:thresholds} (at $t/t^*=150$) with that shown in \cite{DaSilva2008} (at $t/t^*=27$) one observes that Fig. \ref{fig:thresholds} has many more "holes".
This suggests that an enstrophy threshold relative to the levels inside the turbulent core or taking the self-similarity into account is preferable to ensure a consistent interface detection over extended periods of time.

\begin{figure}
 \centering
 \includegraphics{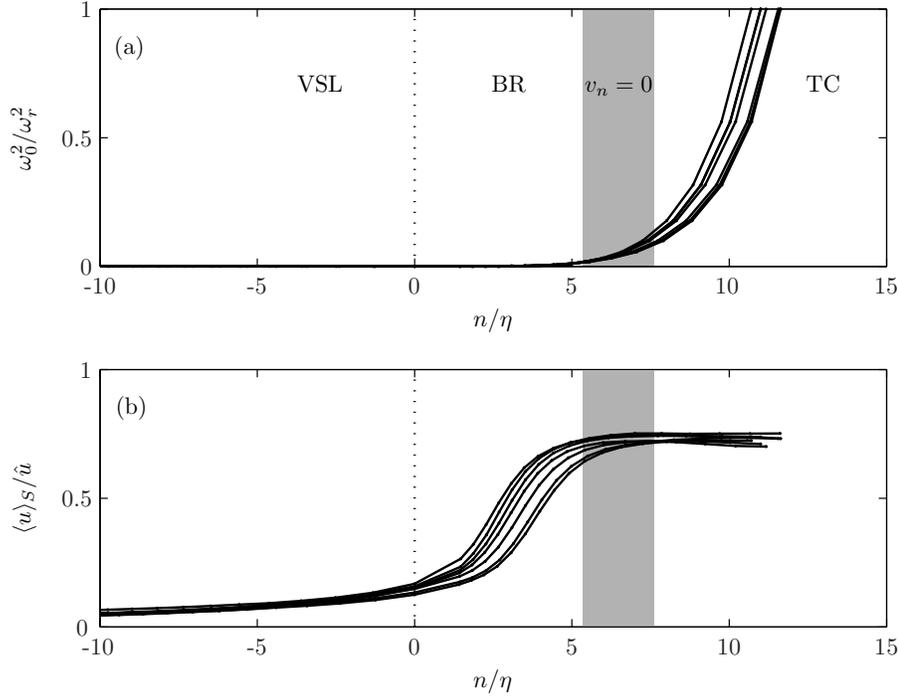}
 \caption{\label{fig:jump} (a) enstrophy as a function of $n$; b) conditioned streamwise velocity $\avS{u}$.}
\end{figure}

A characteristic feature of the TNTI is that when moving into the turbulent layer from the detected interface, the enstrophy quickly increases, peaks and then saturates at a fixed value slightly lower than the peak \citep{Bisset2002, Westerweel2005, DaSilva2008}.
In Fig. \ref{fig:jump}(a), the enstrophy $\omega_0^2/\omega_r^2$ is plotted as a function of $n$, for the period $150 < t/t^* < 300$.
In principle, this plot shows the same information as Fig.\ \ref{fig:VSLthickness}(a) but the axes are now linear.
The viscous superlayer (VSL), buffer region (BR) and turbulent core (TC) are shown for convenience; the region for which $v_n=0$ moves outward as time progresses and is denoted by a grey area.
Consistent with \cite{Bisset2002,DaSilva2008, Mathew2002}, the enstrophy can be seen to increase very rapidly in the transition from the BR to the TC.
There is no plateau in enstrophy because the statistics presented here were obtained by conditioning on the enstrophy, not on the distance to the interface.

Another characteristic feature of the TNTI is a rapid change in the  streamwise momentum \citep{Mathew2002, Bisset2002, Westerweel2005, DaSilva2008}.
In Fig.\ \ref{fig:jump}(b), the conditioned streamwise velocity $\avS{u}/\hat{u}$ is plotted as a function of $n$.
It is clear that the chosen origin is not ideal as there is no full data collapse, but it is evident that within the BR, $\avS{u}$ increases rapidly.
From figure \ref{fig:jump}, it can be concluded that the TNTI likely comprises part of the BR and part of the TC.
The TNTI does not contain the VSL and we thus conclude that the conjecture made by \cite{Bisset2002} is correct - the VSL forms the outer boundary of the TNTI.
The two are dynamically different and will consequently behave differently.

\subsection{What makes the interface propagate?}

One may speculate on the mechanism by which the turbulence boundary moves outward. \cite{Mathew2002} present an argument on how nibbling by  small-scale eddies on the Kolmogorov microscale is compatible with an inviscid macroscale entrainment process using the fractal properties of the TNTI.
Conversely, \cite{Hunt2008} present an argument that larger scales of the order of the thickness of the interfacial shear layer are responsible for the net movement of the interface using a conceptual model of an eddy impinging onto a shear layer.
This paper shows that the VSL is a very thin and almost one-dimensional layer governed by molecular processes that envelopes the turbulence.
Below we argue that the VSL is maintained by a balance between molecular processes and the creation of a large surface area by motions over a range of scales, thereby creating a dynamic equilibrium with associated entrainment velocity of $O(u_\eta)$.

Indeed, the net mean motion of the VSL can be inferred from the integral  scale entrainment flux $Q_e = \alpha_0 \hat{u}$, which can be expressed in turbulence quantities (as characterized by $e^{1/2}$) by $Q_e \sim A e^{1/2}$ because of self-similarity (Fig. \ref{fig:SSjet}).
Due to the fractal nature of the interface, the surface area on the Kolmogorov scale $\eta$ is given by $S \sim A (\eta/b)^{2-D_f}$, where $D_f = 7/3$ is the fractal dimension \citep{Sreenivasan1991}.
Consequently, 
\begin{equation}
  Q_e \sim S \left( \frac{\eta}{b} \right)^{1/3} e^{1/2} = S (\eta \varepsilon)^{1/3} = S u_\eta,
\end{equation}
where an equilibrium across scales of the form $\varepsilon \sim e^{3/2}/b$ was assumed in the second step. The argument above suggests in the VSL $v_n \sim u_\eta$, consistent with Corrsin's argument, previous \citep{Holzner2008,Holzner2011,Wolf2012} and the present work. The fractal geometry argument suggests that $S$ is convoluted over a range of scales from large to small. Referring back to Fig. \ref{fig:vndec_vn}(c,d), we show that on average the curvature of the VSL is low. This would exclude vigorous mixing on the Kolmogorov scales as a dominant process as this would create very strongly curved surfaces. Hence, it seems more plausible that motions on larger scales are more significant in close proximity to the VSL, as the vorticity is oriented and stretched tangentially to the VSL remaining correlated over larger distances, while diffusing viscously in normal direction along which it decays very sharply \citep{Holzner2011}. The surface area $S$ will adjust to the molecular processes governing the VSL by stretching until the product $v_n S$ balances the inviscid entrainment volume flux $Q_e$.
}

\section{Concluding remarks}
\label{sec:conclusions}
\rev{In this paper we studied the structure of the turbulence boundary of a temporal plane jet.}
We find convincing evidence for the existence of the viscous superlayer (VSL). Consistent with earlier work, we find that inertial processes are negligible in the VSL. The VSL is discernible for nearly twenty orders of magnitude in enstrophy threshold. \rev{Taking into account the entire range of thresholds explored, one may attribute a thickness on the order of 15$\eta$ or $\lambda$ to the VSL, which is present for roughly $\omega_0^2/\omega_r^2 < 10^{-5}$.
However, it should be emphasized that there is no natural threshold to define the boundary between the irrotational fluid outside the jet and the VSL. The lower this threshold is, the thicker the VSL will seem. \cite{Holzner2011} quantify the VSL thickness by defining it as $\delta_{\textnormal{VSL}} = (d \omega^2 / dn) / (d^2 \omega^2 / dn^2)$, which in the light of the conceptual model for the VSL, Eq. \eqref{eq:om2_ansol}, corresponds to the e-folding length which is $\eta$.}
The simple theoretical model derived in \S \ref{sec:vnprediction} is in good agreement with the data and shows that the contribution of the viscous transport term amounts to $2 v_n$ and the viscous destruction term to $-v_n$.

\rev{The simulations support the classical assumption that global entrainment is independent of molecular processes, which was clear from the fact that the entrainment flux $Q_e = v_n S$ was practically independent of the threshold in the VSL.
The local entrainment velocity $v_n$ was of the same order of magnitude as $u_\eta$, although there was a dependence of the value of $v_n$ on the enstrophy threshold.}
This suggests that Corrsin's dimensional arguments may need modification for the moderate Reynolds number under consideration here.
\rev{Indeed, in the VSL} both viscous transport velocity and surface area adjust to the imposed global rate so that the \rev{product $v_n S$ is practically independent of $\omega_0^2$}. As the VSL becomes less contorted when moving out of the turbulent region $v_n$ needs to become larger to maintain a constant entrainment flux.
\rev{The small dependence of $Q_e$ on $\omega_0^2$ could be explained by taking into account that the average position $\avS{y}/b$ of the interface has a non-negligible variation across the turbulence boundary.}

Three regions were observed for the flow under consideration. For roughly $\omega_0^2/\omega_r^2 < 10^{-5}$, we observed the VSL which is characterised by a propagation velocity only depending on viscous processes.
The turbulent core region (TC) was categorised as that region of the flow for which $v_n > 0$, i.e.\ regions that become smaller in time and that are responsible for the $\av{\fl{\omega}\fl{\omega}} \propto t^{-2}$ behaviour.
The enstrophy threshold at which this region was observed was a decreasing function of time.
\rev{In between the VSL and the TC there is a buffer region (BR) providing a smooth transition from the VSL and the TC.
The buffer region shows many similarities with the TNTI, although the TNTI seems to extend into the TC. \rev{The core region is on average about $10 \eta$ or $0.5 \lambda$ away from the VSL, this defines the thickness of the BR}.
The current work suggests that the VSL forms the outer boundary of the TNTI, confirming the conjecture made by \citet{Bisset2002}.
Further work, particularly a study of the $\Rey$ dependence, is necessary to clarify the similarities and differences between the BR and TNTI.}

\rev{Our analysis spans 24 orders of magnitude in enstrophy levels, generalizing previous approaches that are mostly based on a single threshold value. This systematic approach hence allowed us to overcome the degree of arbitrariness associated with the choice of a single threshold. Other approaches (e.g. automatic approaches) to choose an appropriate threshold for the identification of the turbulence boundary may be possible. As long as such a method is not in place, we advocate here that the analysis of a complete span of thresholds is advisable, especially because it allows separating between physically distinct regions that constitute the turbulence boundary, namely TC, BR and VSL.
}

\section*{Acknowledgements}

The computational resources for this work were provided by the Imperial College HPC facilities. Both authors benefited greatly from an inspiring MULTIFLOW workshop on the Turbulent/Non-turbulent Interface which took place in October 2012 in Madrid. M.~H. acknowledges financial support from the Swiss National Science Foundation (SNF) under grant number PBEZP2-127831 and from the European Commission under the Marie Curie Intra-European Fellowship, project number 272886.

\end{document}